\documentclass[epj]{svjour}
\usepackage{graphics}
\usepackage{amsmath}
\begin{document}
\title{A systematic approach to Covariance matrix formulation in charged particle activation experiments}
\author{Tanmoy~Bar\thanks{\email{amitanmoybar@gmail.com}}}
\offprints{Tanmoy}          
\institute{Inter University Accelerator Centre, Aruna Asaf Ali Marg, Vasant Kunj, New Delhi-110067, India}
\date{Received: date / Revised version: date}
%
\abstract{
This work presents a detailed covariance and correlation matrix analysis for experimentally measured cross sections obtained using the activation technique. Both statistical and systematic contributions to the covariance matrix were explicitly calculated using sensitivity coefficients. The detector efficiency was determined by refitting standard source data with an exponential function, and the associated covariance matrix of the fitted parameters was propagated to estimate the uncertainty in efficiency at the relevant $\gamma$-ray energy. The cross sections and the corresponding experimental parameters, such as beam flux, target thickness, $\gamma$-ray intensity, and decay corrections, were taken from previously published measurements and are used here for the purpose of illustrating the covariance formalism. The resulting covariance and correlation matrices provide a comprehensive representation of uncertainties and their interdependencies. This formalism demonstrates the importance of including correlated uncertainties for reliable interpretation and comparison of experimental cross section data.
} 
\maketitle
\section{Introduction}
\label{intro}
The cross section measurements of nuclear reactions using the offline activation method have been utilized for many decades and find applications in various fields such as radioisotope production, reaction dosimetry, and related areas~\cite{qaim2012present,upadhyay2025production,carlson2009international,capote2018iaea,qaim2017nuclear}. This technique has been widely employed in nuclear reaction studies involving different projectiles, including neutrons, photons, and charged particles, for several decades~\cite{choudhary2022measurement,michel1997cross,de2001evaluation,tarkanyi2003excitation}. In this article, the activation study with the charged particle projectile is discussed. For the last few decades the charged particle activation method (CPAM) has been used to a great extend for the nuclear astrophysics cross section measurements. In nuclear astrophysics studies determination of reaction rate is one of the most crucial aspect. In order to calculate the reaction rate of the nuclear reaction at astrophysical energies, only the total cross section is required. The activation technique is one of the most powerful tool to get the information about the total cross section in case of the radioactive residual nuclei of the reaction.
Accurate cross section measurements obtained using the CPAM play a central role in many domain such as nuclear physics, nuclear astrophysics, medical isotope production, and applied nuclear technologies. In particular, this method can provide reliable cross sections at low and intermediate energies (important for many nuclear astrophysics studies) where in-beam technique is extremely challenging due to extreme low cross section values ($\sim$pb-nb) in presence of in-beam and cosmic background.

Because these data are frequently incorporated into the evaluated nuclear data libraries and used for model validation (eg. statistical model calculation) their uncertainties are very crucial to be presented in a rigorous and transparent manner. Traditionally, the experimental uncertainties have been presented as total uncertainties obtained from the quadratic summation of statistical and systematic contributions. while this method provides an estimation of overall error magnitude, but greatly ignores the correlation between measurements performed at different energies. This correlation becomes even more significant in measurements involving low cross-section values and a steep energy dependence, a trend commonly observed in low-energy nuclear physics~\cite{gyurky2019activation,bar2024measurement,basak2024experimental,saha2025proton,bar2026excitation,kappeler2011s}.

In activation experiments, several uncertainty components, such as detector efficiency, beam flux, target thickness, decay data etc. are the common parameters to all energy points and therefore introduce non-negligible correlations. Overlooking these may lead to bias model estimations and inconsistent nuclear data evaluations. The proper treatment of these correlation uncertainties requires the construction of covariance matrix. In this article, the Jacobian based propagation formulation~\cite{Bevington2003,otuka2017uncertainty,Cowan1998} is discussed for covariance matrix construction.
In the field of nuclear data, the importance of covariance information has long been emphasized. As Donald L. Smith \textit{et al.}~\cite{smith1987generation} discussed that the covariance matrices are indispensable for reliable data evaluation, meaningful comparison between experimental data sets. Evaluated nuclear data libraries distributed through IAEA OECD nuclear energy agency~\cite{chadwick2011endf} explicitly require covariance information to support uncertainty propagation in reactor design, astrophysical reaction rate calculations, and applied modeling. Without detailed covariance elements, the statistical weight of experimental data cannot be correctly assigned in global model fits, and the impact of these systematic normalization uncertainties remains obscured.

For charged particle activation measurements in particular, covariance effects are often dominated by normalization parameters (eg. beam flux, detector efficiency, decay information etc.) that are fully correlated across energies. The explicit construction and reporting of covariance matrices therefore enhances the reliability and long term usability of the measured quantities.

The present work discusses a systematic, step-by- step guideline for constructing covariance matrices exclusively for charged particle activation cross section measurements. The emphasis is placed on the identification of all possible uncertainty sources, formulation of sensitivity  coefficients, separation of correlated and uncorrelated components and realistic implementation within the Jacobian formalism . The main objective of this article is to offer a transparent, easy and reproducible workflow that can be followed in future charged particle activation measurements for covariance matrix construction and nuclear data reporting.

\section{Theoretical background}
\label{sec:2}
\subsection{Basics of charged particle activation measurements}
\label{sec:2.1}
The nuclear reaction which involves radioactive reaction products of sufficient half life, the cross section can be measured using offline measurements of the number of produced isotopes.This method, where applicable, can offer substantial advantages over traditional in-beam $\gamma$-ray or charged-particle detection techniques, particularly in cases where in-beam background (mainly $\gamma$-rays) hampers detection and precise beam-spot definition is not required.

In this technique, a target is irradiated with the beam (such as neutron, photon, proton etc.), producing radioactive nuclei through reaction. After this, the induced activity of the targets is measured offline using high resolution gamma-ray detectors (charged particle detectors in case of charge particle decay~\cite{somorjai1997experimental}). The cross section is then determined from the measured activity by accounting for beam current, detector efficiency, target thickness etc. This technique entirely relies on the post irradiation decay measurements rather than in-beam detection, it offers high sensitivity, excellent signal to noise ratio, and reliable normalization, making it extremely suitable for low cross section measurements.

In this article, our discussion is restricted to activation measurements induced by charged particles and detected via gamma-ray spectroscopy.
\subsection{Cross section formalism}
\label{sec:2.2}
In activation, the cross section measurements are done by measuring the activity ($A$) of the bombarded targets using,
\begin{equation}
A=\frac{\lambda C_\gamma}{\varepsilon I_\gamma~e^{-\lambda t_c}(1-e^{-\lambda t_m)}}
\label{eq2.1}
\end{equation}
where $\lambda~(=\frac{0.693}{T_{1/2}})$ is the decay constant, $\varepsilon$ stands for detector efficiency, $C_\gamma$ is the count under the peak, $I_\gamma$ is $\gamma-$ray emission intensity, $t_c$ is the time between end of irradiation and start of counting, and $t_m$ is the counting time.

Then the cross section is determined by,
\begin{align}
\sigma &= \frac{A}{\phi_b N_t (1-e^{-\lambda t_{irr}})} \\
 &= \frac{\lambda C_\gamma}{\phi_b~N_t~I_\gamma~\varepsilon~(1-e^{-\lambda t_{irr}}) (e^{-\lambda t_c} - e^{-\lambda(t_c + t_m)})}
\label{eq2.2}
\end{align}
where $\phi_b$ denotes proton flux (1/s), $N_t$ is the target atoms/cm$^2$ and $t_{irr}$ is the irradiation time.
The cross section formula consists of all the parameters which contributes towards the final uncertainty in the measurements. Apart from parameter uncertainties, efficiency estimation of the detector also possesses uncertainty due to the curve fitting. The procedure to determine the efficiency uncertainty is discussed in later section.

In many experiments, the projectile flux is determined using a monitor reaction of well documented cross section. It is determined rearranging Eq.~\ref{eq2.1},
\begin{equation}
\phi_b = \frac{A}{\sigma_M N_t (1-e^{-\lambda t_{irr}})} 
\label{eq2.3}
\end{equation}
where $\sigma_M$ is the well known monitor reaction cross section. In such cases, the uncertainty  in the $\sigma_M$ contributes to the beam flux determination.
\section{Sources of different uncertainties}
\label{sec:3}
The accurate determination of any measurement requires a comprehensive identification and classification of all uncertainty components. The total uncertainty associated with the measurement arises from both statistical and systematic effects related to the measurement technique or method. In a charged particle activation cross section measurement gamma counting, detector calibration etc. also contributes. A clear distinction between these contributions is essential for proper representation of total uncertainty.

\subsection{Statistical uncertainties}
\label{sec:3.1}
The statistical uncertainties originate from the inherent random nature of radioactive decay and counting method. In activation measurements the dominant statistical contribution arises from the counting statistics of the observed gamma ray peak (area under the gamma peak). The relative statistical uncertainty is inversely proportional to the square root of the number of obtained counts.

\subsection{Systematic uncertainties}
\label{sec:3.2}
The systematic component of the uncertainty stem from experimental parameters that affect the absolute normalization of the measured cross section. The typical sources of this include detector efficiency, beam current determination, target thickness, decay data ($\gamma-$ray emission intensity and half life), and timing correction for irradiation, cooling and measuring intervals.

\subsection{Correlated and uncorrelated contributions}
\label{sec:3.3}
In addition to their statistical or systematic nature, uncertainty components must be classified according to their correlation structure.

Uncorrelated uncertainties affect each measurement independently. In CPAM, the counting statistics represent the major uncorrelated contribution, since each irradiation and activity measurements are statistically independent.

Correlated uncertainties arises from parameters common to multiple energy points, such as detector efficiency, beam intensity, adopted decay data etc. These quantities act as a common/global normalization factors; any variation in these parameters coherently affect all the cross sections.

In some cases, few parameters such as target areal density may act as a uncorrelated, fully correlated or partially correlated. When different targets are used for different energy data points, the correlation is broken as it was in the case of single (same) target use for all the energies. In an experiment, the same target can be used for multiple Irradiation depending on the half-life of the reaction product of interest. Even when different targets are used, the method employed for determining the target thickness can introduce a partial correlation between the measurements. In the present work, we consider the case where each irradiation at a given energy is performed using a separate target; therefore, the contributions to different cross section measurements are assumed to be uncorrelated.

A realistic treatment of experimental uncertainties therefore requires explicit identification of their correlation structure, whether fully or partially correlated or independent, in order to ensure consistent uncertainty propagation and reliable comparison of cross section data across different energies. The formal incorporation of these correlation properties into a unified mathematical framework is discussed in the following section, where construction of a covariance matrix is discussed in details.

\section{Mathematical framework for Covariance matrix construction}
\label{sec:4}
In experimental measurements, the determination of cross section is influenced not only by the individual measurement uncertainties but also by the correlations arising from common normalisation parameters.

From Eq.~\ref{eq2.2}, it can be followed that at a irradiation energy, $E_i$ the cross section can be written in a general form as,
\begin{align}
\sigma_i&=\xi(x_k) \\
x_k &\equiv \bigl \{ C_i, \varepsilon_i, I_\gamma, \phi_{b,i}, N_{t,i}, \lambda, t_{irr,i}, t_{c,i}, t_{m,i} \bigr \}.
\label{eq4.1}
\end{align}
Each cross section therefore depends on multiple parameters, some are energy dependent and some of which are common to all energies.

\subsection{Sensitivity coefficient and Jacobian matrix}
\label{sec:4.1}
The propagation of total uncertainty requires evaluation of the sensitivity of the cross section to each parameter. These are expressed through the partial derivatives,
\begin{equation}
J_{ik}=\frac{\partial \sigma_i}{\partial x_k}
\label{eq4.2}
\end{equation}
where $J_{ik}$ is the element of Jacobian matrix.

For activation cross section determination formula (Eq.~\ref{eq2.2}), these elements become
\begin{align}
&\frac{\partial\sigma_i}{\partial C_i}= \frac{\sigma_i}{C_i}; \hspace{0.25cm} &\frac{\partial\sigma_i}{\partial\phi_{b,i}} &= -\frac{\sigma_i}{\phi_{b,i}}\nonumber \\ 
&\frac{\partial\sigma_i}{\partial N_{t,i}}= -\frac{\sigma_i}{N_{t,i}}; &\frac{\partial\sigma_i}{\partial\varepsilon_i} &= -\frac{\sigma_i}{\varepsilon_i}\nonumber \\ 
&\frac{\partial\sigma_i}{\partial I_\gamma}= -\frac{\sigma_i}{I_\gamma}
\label{4.3}
\end{align}
The sensitivity coefficient of the decay constant and time parameters differs from other parameters because the decay constant enters the activation equation through exponential decay corrections. While most parameters appear as simple multiplicative factors (providing constant sensitivities of $\pm$1), the decay constant impacts the irradiation, cooling, and counting time terms, leading to a more complex sensitivity expression that depends on the time factors. Similarly, the sensitivity coefficients of the time parameters are not constant and depend explicitly on the value of the decay constant. The expressions for the partial derivatives are given as follows.
\begin{align}
&\frac{\partial\sigma_i}{\partial \lambda}= \frac{\sigma_i}{\lambda} S_{\lambda,i} \label{4.4a}\\
&\text{where,}\nonumber \\ 
&S_{\lambda,i} = \Biggl[1-\frac{\lambda t_{irr,i}~e^{-\lambda t_{irr,i}}}{1-e^{-\lambda t_{irr,i}} } +\lambda t_{c,i}-\frac{\lambda t_{m,i}~e^{-\lambda t_{m,i}}}{1-e^{-\lambda t_{m,i}} } \Biggl]
\label{4.4}
\end{align}
\begin{align}
&\frac{\partial\sigma_i}{\partial t_{irr,i}}= \sigma_i \frac{\lambda e^{-\lambda t_{irr,i}}}{1-e^{-\lambda t_{irr,i}}}; \hspace{0.5cm} \frac{\partial\sigma_i}{\partial t_{c,i}}= \lambda \sigma_i \nonumber  \\
&\frac{\partial\sigma_i}{\partial t_{m,i}}= \sigma_i \frac{\lambda e^{-\lambda t_{c,i}}}{1-e^{-\lambda t_{c,i}}}.
\label{4.5}
\end{align}
Although the contributions from the decay constant and timing parameters to the overall covariance are generally small, the contributions specially from the timing parameters are often neglected in the calculation.

For an ideal condition when $t_{irr,i}$ is sufficiently large, the second term in Eq.~\ref{4.4} vanishes. Similarly, for a large counting time, the fourth term can also be treated as negligible. If the cooling time is kept sufficiently small, $S_\lambda$ approaches unity. Under these conditions, the matrix element associated with the decay constant takes a form similar to that of the other parameters. But in practical scenarios, the uncertainty in $\lambda$ is relatively small $(\frac{\Delta\lambda}{\lambda}=\frac{\Delta t_{1/2}}{t_{1/2}})$, and in most cases $\lvert S_\lambda \rvert < 1$. This makes the contribution of the half life to the overall uncertainty small, and it can often be neglected.

\subsection{Uncertainty in the detector efficiency}
\label{sec:4.2}
The detector efficiency for a specific $\gamma-$ray is determined from the fitted curve of efficiency versus $\gamma-$ray energy, obtained using calibration sources that emit multiple $\gamma-$ray of known energies, such as $^{152}$Eu, $^{60}$Co etc.
The detection efficiency is calculated using Eq.~\ref{eq2.1}, which reduces to the following expression when the half-life is much larger than the measuring time ($\lambda t_m<<1$),
\begin{equation}
\varepsilon_i=\frac{C_i K_i}{A_0 I_\gamma~e^{-\lambda t_{c,i}}t_{m,i}}
\label{4.6}
\end{equation}
where $A_0$ is the activity of the calibration source at the time of its production, and the factor $K_i$ is introduced to account for the coincidence summing effect. The uncertainty in $\varepsilon_i$ is calculated using the quadratic sum of the individual uncertainties of $C_i$, $I_\gamma$, and $A_0$, while the uncertainties due to $t_{c,i}$ and $t_{m,i}$ are considered negligibly small.

These data points are then fitted using the relation
\begin{equation}
\varepsilon(E_\gamma)=\varepsilon_0 e^{-E_i/E_0}+\varepsilon_c
\label{4.7}
\end{equation}
and parameters $\varepsilon_0$, $E_0$ and $\varepsilon_c$ are extracted along with the corresponding parameter covariance matrix. Most commonly used fitting programs, such as Cern ROOT package~\cite{brun1997root} or OriginLab~\cite{OriginManual}, provide this covariance matrix as part of the least-squares fitting procedure.

The obtained covariance matrix can then be used to determine the efficiency and the corresponding uncertainty at any $\gamma-$ray energy within the fitted region. The uncertainty in $\varepsilon(E_\gamma)$ (Eq.~\ref{4.7}) is expressed by $u_\varepsilon(E_\gamma)$ as 
\begin{equation}
u_\varepsilon^2(E_\gamma)=\sum_{i,j}(\frac{\partial \varepsilon}{\partial p_i})~\mathcal{V}_{ij}~(\frac{\partial \varepsilon}{\partial p_j})
\label{4.8}
\end{equation}
where $p_i$ and $p_j$ represent the fitting parameters $\varepsilon_0$, $E_0$ and $\varepsilon_c$, and $\mathcal{V}_{ij}$ is the covariance matrix element between the $i^{th}$ and $j^{th}$ parameters obtained from curve fitting.
From Eq.~\ref{4.7} it can be followed as,
\begin{align}
&\frac{\partial \varepsilon}{\partial \varepsilon_0}= e^{-E_i/E_0} \\
&\frac{\partial \varepsilon}{\partial E_0}= \varepsilon_0~(\frac{E_i}{E_0^2})~e^{-E_i/E_0} \\
&\frac{\partial \varepsilon}{\partial \varepsilon_c}= 1.
\label{4.9}
\end{align}
Eq.~\ref{4.8} can be expressed in matrix form as 
\begin{equation}
u_\varepsilon^2(E_\gamma)=\textbf{g}^T\mathcal{V}~\textbf{g}
\label{4.10}
\end{equation}
where $\mathbf{g}$ is the vector of partial derivatives of $\varepsilon(E_\gamma)$ with respect to the fitting parameters, arranged consistently with the ordering of the covariance matrix $\mathcal{V}$. Accordingly, the transpose of the vector $\mathbf{g}$ is expressed as
\begin{eqnarray}
\textbf{g}^T = 
\left(
\begin{array}{ccc}
\frac{\partial \varepsilon}{\partial \varepsilon_c} &
\frac{\partial \varepsilon}{\partial \varepsilon_0} &
\frac{\partial \varepsilon}{\partial E_0}
\end{array}
\right)
\label{4.11}
\end{eqnarray}
with the covariance matrix given by
\begin{equation}
\mathcal{V} =
\begin{pmatrix}
\text{cov}(\varepsilon_c,\varepsilon_c) &~ \text{cov}(\varepsilon_c,\varepsilon_0) &~ \text{cov}(\varepsilon_c,E_0) \\
\text{cov}(\varepsilon_0,\varepsilon_c) &~ \text{cov}(\varepsilon_0,\varepsilon_0) &~ \text{cov}(\varepsilon_0,E_0) \\
\text{cov}(E_0,\varepsilon_c) &~ \text{cov}(E_0,\varepsilon_0) &~ \text{cov}(E_0,E_0)
\end{pmatrix}.
\end{equation}

\section{Construction of covariance matrix}
\label{sec:5}
The covariance matrix ($V_\sigma$) of the measured cross sections can be expressed as the sum of two terms as follows:
\begin{equation}
V_\sigma=V^{stat}+V^{sys}
\label{5.1}
\end{equation}
where $V^{stat}$ and $V^{sys}$ represent the statistical and systematic contributions arising from different parameters, as discussed in Section~\ref{sec:3}. The compact Jacobian form of covariance matrix expression is
\begin{equation}
V_\sigma= J \eta J^T
\label{5.2}
\end{equation}
where $J$ is the Jacobian matrix having elements $\frac{\partial \sigma_i}{\partial x_k}$ (see Section~\ref{sec:4.1}); $\eta$ is the covariance matrix of these parameters expressed as
\begin{equation}
\eta_{kl}=\mathrm{cov}(x_k, x_l)
\label{5.3}
\end{equation}
The ($i,j$) element of $V_\sigma$ (relating $i^{th}$ and $j^{th}$ measurement) is
\begin{align}
V_{ij}&=\sum_{k=1}^m \sum_{l=1}^m J_{ik}~\eta_{kl} J_{jl} \\
&=\sum_{k=1}^m \sum_{l=1}^m \frac{\partial \sigma_i}{\partial x_k}~\frac{\partial \sigma_j}{\partial x_l}~\mathrm{cov}(x_k, x_l)
\label{5.4}
\end{align}
where $m$ is the number of parameters in the equation. Assuming that the input parameters are mutually uncorrelated, the covariance between parameters satisfies $\eta_{kl}=0$ for $k \neq l$, while the diagonal elements are given by $\eta_{kk}=(\Delta x_k)^2$. Under this assumption, the systematic component of Eq.~\ref{5.4} reduces to
\begin{equation}
V_{ij}^{sys}=\sum_{k=1}^m  \frac{\partial \sigma_i}{\partial x_k}~\frac{\partial \sigma_j}{\partial x_k}~(\Delta x_k)^2
\label{5.5}
\end{equation}
where $\Delta x_k$ denotes the standard uncertainty associated with the parameter $x_k$
The statistical component of the covariance matrix is expressed as
\begin{equation}
V_{ij}^{stat} =
\begin{cases}
(\Delta \sigma_i^{stat})^2, & i = j \\
0, & i \neq j
\end{cases}
\label{5.6}
\end{equation}
reflecting the absence of correlations between statistically independent measurements.

Substituting Eqs.~\ref{4.3} and~\ref{4.4a} into Eq.~\ref{5.5}, the systematic contributions can be written in compact form as
\begin{align}
V_{ij}^{sys}(x_p)= \sigma_i \sigma_j~(\frac{\Delta x_p}{x_p})^2  \label{5.7a}\\
V_{ij}^{sys}(\lambda)= \sigma_i \sigma_j~\left| S_{\lambda i} \right|\left| S_{\lambda j} \right|(\frac{\Delta \lambda}{\lambda})^2
\label{5.7}
\end{align}
where $x_p \in \left\{C, \phi_b, N_t, \varepsilon, I_\gamma \right\}$.

The statistical uncertainty arises from counting statistics, assuming Poisson behavior of the measured counts, such that $\Delta C_i = \sqrt{C_i}$. Accordingly, the variance of the cross section due to statistical fluctuations is given by $(\Delta \sigma_i^{stat})^2 =\frac{\sigma_i^2}{C_i}$ leading to the diagonal covariance matrix elements
\begin{equation}
V_{ii}^{stat} = \Bigl(\frac{1}{\phi_{b,i} \, N_{t,i} \, \varepsilon_i \, I_\gamma \, f(\lambda, t)}\Bigl)^2C_i
\label{5.8}
\end{equation}
where $f(\lambda, t)$ represents the combined decay correction factor accounting for irradiation, cooling, and counting intervals.

The total covariance matrix ($V$) is obtained by summing the statistical and systematic contributions, thereby providing a complete description of both uncorrelated and correlated uncertainties in the measured cross sections,
\begin{equation}
V =
\begin{pmatrix}
V_{11} & V_{12} & \cdots & V_{1j} \\
V_{21} & V_{22} & \cdots & V_{2j} \\
\vdots & \vdots & \ddots & \vdots \\
V_{i1} & V_{i2} & \cdots & V_{ij}.
\end{pmatrix}
\label{5.9}
\end{equation}
where $V_{ii}$ is called the variance of $\sigma_i$ and $V_{ij}$ is the covariance between $\sigma_i$ and $\sigma_j$.
\subsection{Correlation matrix}
\label{sec:5.1}
The correlation coefficient ($\rho_{ij}$) between two measurements is obtained by normalizing each covariance element with their standard deviations (=$\sqrt{\text{variance}}$),
\begin{equation}
\rho_{ij}=\frac{V_{ij}}{\sqrt{V_{ii}V_{jj}}}
\label{5.10}
\end{equation}
Using the above definition the correlation matrix ($R$) becomes,
\begin{equation}
R =
\begin{pmatrix}
1 & \rho_{12} & \cdots & \rho_{1j} \\
\rho_{2} & 1 & \cdots & \rho_{2j} \\
\vdots & \vdots & \ddots & \vdots \\
\rho_{i1} & \rho_{i2} & \cdots & 1
\end{pmatrix}
\label{5.11}
\end{equation}
where $-1\le \rho_{ij} \le 1$ characterizing the degree of correlation between measurements. A value $\rho_{ij}=+1$ corresponds to full correlation, $\rho_{ij}=-1$ to complete anti-correlation, and zero indicates the absence of correlation. Intermediate values reflect partial correlation or anti-correlation between the variables.
\section{Application to an experimental data}
In order to demonstrate the procedure described in the preceding sections, the experimental crosssection data from Ref.~\cite{bar2024measurement} have been utilized. In the present study, these data are employed solely as an illustrative example to demonstrate the covariance and correlation matrix formalism, which was not included in the original work.
\subsection{Brief experimental details}
A 7 MeV proton beam was employed to measure cross sections at different energies using aluminum degrader foils. The $\mathrm{Sm}_2\mathrm{O}_3$ targets, enriched to 67\% in $^{144}\mathrm{Sm}$, were used to determine the $^{144}\mathrm{Sm}(p,\gamma)$ cross sections. Irradiation at each energy were carried out for duration ranging from several hours to a few days, followed by cooling periods varying from a few hours to several days. After the cooling interval, the irradiated targets were placed in front of an HPGe detector with 40\% relative efficiency for $\gamma$-ray spectroscopy measurements. The characteristic 893.73 keV $\gamma$-ray from $^{145}$Eu, with an intensity of $66 \pm 4$\%, was used for cross section determination. Further details of the experimental setup and measurement procedure are available in the referenced work~\cite{bar2024measurement}. In the present study, only the lowest four data points (2.6 to 4.1 MeV) are considered for analysis.

\subsection{Uncertainty details of input parameters}
The detector efficiency data reported in Ref.~\cite{bar2024measurement} were refitted using Eq.~\ref{4.7}, and the corresponding covariance matrix was obtained. The fitted parameters along with their covariance matrix are presented in Table~\ref{tab:1}. 
\begin{table*}
\caption{Fitted efficiency parameters and the corresponding covariance matrix (lower triangular elements).}
\label{tab:1}
\centering
\footnotesize
\begin{tabular}{ccccc}
\hline\noalign{\smallskip}
Parameter & Fitted value & \multicolumn{3}{c}{Covariance matrix} \\
\noalign{\smallskip}
 &  & $\varepsilon_c$ & $\varepsilon_0$ & $E_0$ \\
\noalign{\smallskip}\hline\noalign{\smallskip}
$\varepsilon_c$ & 0.002205 & $3.78848\times10^{-6}$ &  &  \\
$\varepsilon_0$ & 0.13897  & $1.24724\times10^{-5}$ & $1.62665\times10^{-4}$ &  \\
$E_0$           & 329.427  & $-0.06198$ & $-0.41217$ & 1502.97643 \\
\noalign{\smallskip}\hline
\end{tabular}
\end{table*}
The uncertainty in the detector efficiency ($u_\varepsilon$) is evaluated using Eq.~\ref{4.10} as 
\begin{align}
u_\varepsilon^2&=\mathrm{cov}(\varepsilon_c,\varepsilon_c)+(e^{-E_i/E_0})^2 \mathrm{cov}(\varepsilon_0, \varepsilon_0)\nonumber \\
&+\Bigl(\varepsilon_0(\frac{E_i}{E_0^2}) e^{-E_i/E_0}\Bigl)^2 \mathrm{cov}(E_0,E_0)+2e^{-E_i/E_0} \mathrm{cov}(\varepsilon_0, \varepsilon_c) \nonumber \\
&+2\varepsilon_0(\frac{E_i}{E_0^2}) e^{-E_i/E_0} \mathrm{cov}(E_0, \varepsilon_c)\nonumber \\
&+2\varepsilon_0(\frac{E_i}{E_0^2}) \Bigl(e^{-E_i/E_0}\Bigl)^2 \mathrm{cov}(E_0, \varepsilon_0).
\label{5.12}
\end{align}
Using the fitted parameters and covariance matrix listed in Table~\ref{tab:1}, the detector efficiency at $E_\gamma = 893.73$~keV is evaluated using Eq.~\ref{4.7} and~\ref{5.12} as
\begin{equation}
\varepsilon_{893.73}~(\%\Delta \varepsilon)=0.03127~(3.59 \%).
\label{5.13}
\end{equation} 

The experimental parameters along with their corresponding uncertainties are listed in Table~\ref{tab:2}.
\begin{table*}[t] 
\centering 
\caption{Experimental parameters used for cross section determination in Ref.~\cite{bar2024measurement}, along with their associated uncertainties.} 
\begin{tabular}{clcccccc} 
\hline 
$E_{\mathrm{lab}}$ & $C_i$($\%\Delta C_i$) & $\phi_i$($\%\Delta \phi_i=5\%$) & $N_{t,i}$($\%\Delta N_{t,i}=40\%$) & $I_\gamma$($\%\Delta I_\gamma$) & $t_{1/2}$($\%\Delta t_{1/2}$) &$\varepsilon$($\%\Delta \varepsilon$)& $\sigma_i$ \\ 
(MeV) &  & (1/s) & (atoms/cm$^2$) & & (days) && ($\mu$b) \\ 
\hline 
4.11 $\pm$ 0.09 & 10636 (0.97$\%$) & 3.86$\times$ 10$^{12}$ & 2.89$\times$ 10$^{17}$ & & &&129.2 \\ 
3.68 $\pm$ 0.10 & 3360 (1.73$\%$)  & 3.20$\times$ 10$^{12}$ & 2.99$\times$ 10$^{17}$ & 66 $\pm$ 4$\%$& 5.93 &0.0313 &30.0 \\ 
3.17 $\pm$ 0.11 & 1104 (3.01$\%$)  & 4.98$\times$ 10$^{12}$ & 3.11$\times$ 10$^{17}$ & & (0.675$\%$)&(3.59 $\%$)&3.11 \\ 
2.59 $\pm$ 0.13 & 258 (6.23$\%$)   & 4.98$\times$ 10$^{12}$ & 3.43$\times$ 10$^{17}$ & & &&0.24 \\ 
\hline 
\end{tabular} 
\label{tab:2} 
\end{table*}
The sensitivity coefficient of the decay constant, $S_{\lambda,i}$ (Eq.~\ref{4.4}), was evaluated using the corresponding irradiation, cooling, and counting times for each energy. The calculated values are presented in Table~\ref{tab:3}.

\begin{table*}[t] 
\centering 
\caption{$S_{\lambda,i}$ for different energies, calculated using $\lambda = 1.35\times10^{-6}$~s$^{-1}$, along with the corresponding irradiation ($t^{\mathrm{irr}}$), cooling ($t^{c}$), and measurement ($t^{m}$) times.} 
\begin{tabular}{ccccc} 
\hline 
$E_{\mathrm{lab}}$ & $t^{irr}$ & $t^c$&$t^m$ &$S_{\lambda,i}$\\ 
(MeV) &  (s) &	(s) &(s)&\\ 
\hline 
4.11	&	73440	&	277373	&	57600	&	-0.54 \\
3.68	&	69653	&	5695	&	66981	&	-0.90 \\
3.17	&	160042	&	198530	&	79200	&	-0.58 \\
2.59	&	160042	&	271166	&	270000	&	-0.36 \\
\hline 
\end{tabular} 
\label{tab:3} 
\end{table*}
Using Eqs.~\ref{5.7a},~\ref{5.7} and~\ref{5.8}, along with the parameters listed in Tables~\ref{tab:2} and~\ref{tab:3}, the systematic and statistical components of the covariance matrix elements are determined. The resulting covariance matrix is presented in Table~\ref{tab:4}. The correlation matrix, obtained using Eq.~\ref{5.10}, is also included in the same table. The cross section uncertainties are given by the square roots of the diagonal elements (variances) of the covariance matrix.

\begin{table*}[!t]
\caption{Cross section values along with the corresponding covariance and correlation matrices.}
\label{tab:4}
\centering
\footnotesize
\begin{tabular}{cc|llll|llll}
\hline\noalign{\smallskip}
Energy (MeV) & Cross section ($\mu$b) & 
\multicolumn{4}{c|}{Covariance matrix} &
\multicolumn{4}{c}{Correlation matrix (\%)} \\
$(E_{\rm lab}\pm1\sigma)$ & $(\sigma \pm \Delta\sigma)$ \\
\noalign{\smallskip}\hline\noalign{\smallskip}

$4.11\pm0.09$ & $129.24\pm52.55$ & 2760.987 & & & &  100 & & &  \\

$3.68\pm0.10$ & $30.00\pm12.20$ & 20.970 & 148.883 & & &  3.270 & 100 &  & \\

$3.17\pm0.11$ & $3.114\pm1.265$ & 2.171  & 0.505 & 1.599 & &  3.266 & 3.272 & 100  & \\

$2.59\pm0.13$ & $0.242\pm0.098$ & 0.167  & 0.039 & 0.004 & 0.010 &  3.263 & 3.266 & 3.263 & 100  \\

\noalign{\smallskip}\hline
\end{tabular}
\end{table*}

\section{Conclusion}
In this work, a covariance and correlation matrix formalism has been applied to activation-based cross section data, incorporating both statistical and systematic uncertainties through sensitivity coefficients. The analysis demonstrates that statistical uncertainties contribute only to the diagonal elements, while systematic effects introduce correlations between data points. The decay constant exhibits a more complex dependence due to its coupling with irradiation, cooling, and counting times, although its overall impact remains modest. The resulting covariance and correlation matrices provide a complete and consistent representation of uncertainties, essential for reliable data interpretation and comparison. In the present study, the beam flux is assumed to be measured using an electron-suppressed Faraday cup; in cases where monitor foils are used, additional uncertainties from monitor reaction need to be considered, which are beyond the scope of this work and can be found in Ref.~\cite{otuka2017uncertainty}.

The methodology presented here can be readily extended to similar activation-based measurements and highlights the importance of incorporating full covariance information in nuclear data analysis.

\section*{Acknowledgements}
The author would like to thank the Inter-University Accelerator Centre (IUAC), New Delhi, for providing financial support through the Research Associate programme. The author is also grateful to the IUAC colleagues Dr. Akhil Jhingan, Dr. K. S. Golda, Dr. Saneesh N., Mr. Mohit Kumar, and Mr. Rishabh Prajapati for their valuable support and for providing a positive working environment. The author also sincerely acknowledges Prof. Chinmay Basu (Saha Institute of Nuclear Physics, Kolkata) for his guidance and helpful discussions.

%
\bibliographystyle{epj}
\bibliography{ref_file}

\end{document}